\documentstyle[12pt]{article}
\newtheorem{lemma}{Lemma}
\newtheorem{theorem}{Theorem}
\newtheorem{p}{Proposition}

\begin{document}

\title{Axial symmetry and conformal Killing vectors}
\author{Marc Mars and Jos\'e M.M. Senovilla, \\
Departament de F\'{\i}sica Fonamental, Universitat de Barcelona, \\
Diagonal 647, 08028 Barcelona, Spain}

\date{23 October 1992}

\maketitle

\begin{abstract}
Axisymmetric spacetimes with a conformal symmetry are studied and it is shown
that, if there is no further conformal symmetry, the axial Killing vector and 
the conformal Killing vector must commute. As a direct consequence, in
conformally stationary and axisymmetric spacetimes, no restriction is made
by assuming that the axial symmetry and the conformal timelike symmetry commute.
Furthermore, we prove that in axisymmetric spacetimes with another symmetry
(such as stationary and axisymmetric or cylindrically symmetric spacetimes)
and a conformal symmetry, the commutator of the axial Killing vector with the
two others mush vanish or else the symmetry is larger than that originally
considered. The results are completely general and do not depend on
Einstein's equations or any particular matter content.
\end{abstract}

\section{Introduction.}

Symmetry is one of the important issues in Physics and, in particular, it has a
remarkable relevance in General Relativity. Most of the known solutions to the
Einstein field equations have been found by assuming the existence of a more or
less restrictive group of isometries acting on spacetime. In fact, the
classification of spacetimes themselves have been carried out with the help of
the isometries they admit (sometimes in conjunction with other invariant
properties such as Petrov type, etc.), which form a Lie group acting on the
manifold. For a standard review of these matters see the exact solution book
\cite{KSMH}. Most of this work has taken into account true isometries only, but
not homothetic isometries and/or conformal ones.  Fortunately, there has been a
renewed interest for these last types of symmetries recently, and some
classifications have been achieved for special matter contents in the spacetime
or particular structures of the conformal Lie group, (see, for instance,
\cite{CT1}, \cite{CT2}, \cite{CC} and \cite{C-AC}), and also some theorems
 relating
conformal Killings with Killing tensors have been found \cite{Ko}.

Non isometric symmetries have a very important application in stationary and
axially symmetric spacetimes, which constitute one of the outstanding
scientifical subjects within General Relativity due to their importance for the
description of the gravitational field of isolated objects. In fact, most {\it
known} stationary and axisymmetric perfect-fluid solutions possess one of these
symmetries. Namely, the Wahlquist solution \cite{W} (and its limit cases found
by Kramer \cite{K1}, \cite{K2}, \cite{S1}) have a non-trivial Killing tensor,
while the solutions presented in Refs.\cite{S2}\cite{S3} have a proper
conformal Killing vector. It seems therefore very interesting to study general
stationary and axisymmetric spacetimes with one proper (or homothetic)
conformal Killing vector. This study was started by Kramer in a series of
papers \cite{K3},
\cite{K4}, \cite{KC}, where the different Bianchi types of the conformal
 Lie group
and associated line-elements were presented. However, in these papers and also
in
\cite{C-AC} no restriction was considered for the Bianchi types and every
 possible
case was taken into consideration. This is not what happens with Bianchi
 types of
true isometries involving an axial Killing, as
is very well known since the results of Carter \cite{C} concerning axial
symmetry appeared.  Thus, for instance, it follows from Carter's paper that in
stationary and axisymmetric spacetimes, the timelike and axial Killing vectors
commute and, furthermore, either there are only those two symmetries or there
is a larger isometry group of at least four dimensions.  Do similar results
hold for conformal Killings in axially symmetric spacetimes? The answer is yes,
as we shall prove in this paper. More precisely, we are able to show that: 1)
In axially symmetric spacetimes with one and only one conformal Killing vector,
the axial Killing and the conformal Killing commute. Therefore, in axially
symmetric spacetimes with a conformal Killing, if they do not commute then
there is an at least three-dimensional conformal group. 2) In axially symmetric
spacetimes with a {\it timelike} conformal Killing vector, there is no
restriction in assuming that the axial Killing and the timelike conformal
Killing commute and also either there are only those two symmetries or else
there is a, at least, four-dimensional conformal group.  3) In axially
symmetric spacetimes with one more (and only one) Killing vector and one and
only one conformal Killing vector, the axial Killing commute with the other
two.  Therefore, we also have 4) in stationary and axially symmetric spacetimes
with one conformal Killing vector, the axial Killing commutes with the two
others. Thus we see that the three-dimensional conformal group of stationary
and axisymmetric spacetimes with one conformal Killing cannot be arbitrary but
rather it can only take one of the few forms in which the axial Killing
commutes with the other two symmetries. It should be noticed that this is a
general result and it does not depend on the particular form of the
energy-momentum tensor or any other thing at all. Obviously, similar results
hold also for cylindrically symmetric spacetimes with one conformal Killing
vector.

The plan of the paper is as follows. In section 2 we define the concept of
spacetime with axial symmetry and derive a series of previously known theorems
in such a situation.  In particular, we also recall a set of very well-known
properties of the axis of symmetry and the axial Killing vector which are
standard but not easily available in the usual literature. Most of the
notations and results presented there will be used repeatedly in the rest of
the paper. Section 3 is devoted to presenting the main theorems in this work,
which concern the combination of axial symmetry with a conformal symmetry.
Namely, we prove that the axial Killing and the conformal Killing must commute
if we want to have no more symmetry in spacetime. We also prove that if the
conformal Killing is not normal to the axis of symmetry then either commutes
with the axial Killing or there is another conformal Killing that does.  A
trivial corollary of this is that there is no loss of generality in assuming
that a timelike conformal Killing vector commutes with the axial Killing. Of
course, all these results are also valid for homothetic and true Killing
vectors and thereby we reobtain, in this last case, the main results already
known for stationary and axisymmetric spacetimes or cylindrically symmetric
spacetimes. Finally, we devote section 4 to the case in which there is another
Killing vector other than the axial Killing vector and also a proper (or
homothetic) conformal Killing vector. We show that in this case the axial
Killing must commute with the other symmetries. Our notation is standard and
every object that appears herein is defined the first time it comes out.

\section{Definitions and basic results.}

A space-time, $V_{4}$, has axial symmetry if there is an effective realization
of the one-dimensional torus $T$ into $V_4$ that is an isometry and such that
its set of fixed points constitutes a two-dimensional surface.  (It can be seen
that given an isometric realization of the one-dimensional torus whose fixed
points form a two-dimensional surface, there is always another realization with
the same properties and also effective). Mathematically, these two conditions
are expressed by:
\begin{enumerate}
     \item There is a map $\tau$
\begin{eqnarray*}
     \tau  :  T \times V_4 &\rightarrow& V_4 \\ 
     (\phi,x) &\rightarrow& \tau(\phi,x)\equiv\tau_{\phi}(x) 
\end{eqnarray*} 
      which is a realization of the Lie group $T$ where each $\tau_{\phi}$
       is an
      isometry of $V_4$.
      
     \item The set of fixed points of $\tau$, that is to say
     \[W_{2} \equiv \ \symbol{123} \ x \in V_4 \ ; \ \tau_{\phi} (x)
      = x \hspace{0.5cm} \forall \phi  \in T \ \symbol{125} \]
      is a two-dimensional surface in $V_{4}$. From now on, we shall often
      refer to $W_2$ as the {\it axis of symmetry}.
      
\end{enumerate}
The Killing vector field that $\tau$ defines will be called $\vec{\sigma}$
throughout this paper.
We do not assume anything about the spacelike, timelike or null character of
$\vec{\sigma}$ in the spacetime, because this character can change from
one region to another. Despite this fact, it can be proven that 
$\vec{\sigma}$ is always spacelike in a neighbourhood of the axis $W_2$, as
we shall presently see.
At any point $Q \in W_{2}$ we denote by $ L_Q $ the tangent plane of $ W_2 $
at $Q$. Obviously, we have \( L_Q \subset T_Q(V_4) \) where $T_Q(V_4)$ is the
tangent space at $Q$. As a trivial consequence of the above, the differential
 map
\begin{eqnarray*}
d \tau_{\phi} \mid_{x} : T_{x}(V_{4}) &\rightarrow& T_{\tau_{\phi}(x)}(V_{4})
 \\
\vec{V} &\rightarrow& d \tau_{\phi} \mid_{x}(\vec{V})
\end{eqnarray*}
is a linear isometry between tangent spaces, where we denote by $\mid_{x}$ the
restriction to the point $x$. In addition, if $Q \in W_{2}$
then $d \tau_{\phi} \mid_{Q}$ is an isometric automorphism of $T_{Q}(V_{4})$.

There is an easy characterization of the  vectors tangent to the axis of
symmetry in a fixed point, which is intuitively obvious having in mind that
$d \tau_{\phi}$ is the linearization of the isometry group. This
characterization is written down in the following Lemma (see e.g. \cite{CHUS}):

\begin{lemma}
  For points $Q$ on the axis, \hspace{0.2cm} 
  $\vec{V} \in L_Q$ if and only if $d\tau_{\phi} \mid_Q
  \vec{V} = \vec{V}$ for every $\phi \in T$.
\end{lemma}
  In other words, the vectors in $L_Q$ ( $Q \in W_2$) are those invariant
by $\symbol{123}\tau_{\phi}\symbol{125}$. 

For points on the axis, we define $P_Q$ as the linear subspace of $T_Q(V_4)$
orthogonal to $L_Q$. Being $d \tau_{\phi} \mid_Q $ an isometric automorphism
of $T_Q(V_4)$ it leaves $P_Q$ invariant. The following general result about 
representations of the one-dimensional
torus allows us to prove that the axis of symmetry is a timelike surface.

\begin{lemma}
 Let $\cal{V}$ be a two-dimensional vector space with a non-vanishing metric.
 If there is a (continous) representation $\{R_{\phi}\}$ of the Lie group
 $T$ on $\cal{V}$ which is an isometry and also there is a $\phi_0 \in T$
  such that
 $R_{\phi_{0}} \not= Id \mid_{\cal{V}}$, then the metric in $\cal{V}$ must
 be positive
 definite.
\end{lemma}
We have already shown that $P_Q$ is invariant under $d\tau_{\phi}$, which 
is a representation of $T$. If we had  
$d\tau_{\phi} \mid_{P_Q} = Id \mid_{P_Q}$ for all $\phi \in T$ , then by
 Lemma 1 we would have 
$P_Q \subset L_Q$ which is impossible in a linear space with a lorentzian 
metric. Then, we can apply the previous Lemma to deduce that $P_Q$ must be 
positive definite. Therefore $L_Q$ is a two-dimensional lorentzian subspace 
and $P_Q \cap L_Q = \{\vec{0}\}$. Thus we have proven (see \cite{C} or
\cite{CHUS}):

 \begin{theorem}
 The axis of symmetry, $W_2$, is time-oriented.
 \end{theorem}

    In what follows, we are going to prove very general results concerning
    the commutators of general vector fields with the axial Killing vector
    field.
    These general results will be essential for the main theorems of sections
    3 and 4 below. To that end, let $\vec{\alpha}$ be any vector field. The
    first 
    Lemma we can show is:
    
    \begin{lemma}
     For $Q \in W_2$, $\vec{\alpha} \mid_Q$ is tangent to the axis if and
      only if
     $[\vec{\alpha},\vec{\sigma}] \mid_Q = \vec{0}$.
    \end{lemma}
    {\it Proof :} Due to Lemma 1 we know that 
    $\vec{\alpha}$ is tangent to the surface of fixed points iff it is
    invariant by $\{\tau_{\phi}\}$ at points on the axis, which is equivalent
    to saying that $\left(\frac{d}{d\phi} d\tau_{\phi}\mid_{\phi=0}\right)
    \vec{\alpha}
    \mid_{Q}=\vec{0}$. But  using the fact that in a coordinate system 
    $\frac{d}{d\phi}(\tau_{\phi})
    _{\mu}^{\beta}\mid_{\phi=0}= \partial_{\mu}\sigma^{\beta}$, this is
    equivalent to
\begin{eqnarray*}
 \alpha^{\mu}\partial_{\mu}\sigma^{\beta}\mid_Q=0. 
\end{eqnarray*} 
Now, given that $\vec{\sigma}\mid_Q=\vec{0}$, we can rewrite this equation as 
$\alpha^{\mu}
    \partial_{\mu}\sigma^{\beta}-\sigma^{\mu}\partial_{\mu}\alpha^{\beta}
    \mid_Q=0$,
that is to say, $[\vec{\alpha},\vec{\sigma}]\mid_Q=\vec{0}$, and the Lemma
 is shown.

\vspace{0.5cm}
    
    Thus, we have a second characterization for vectors in $L_Q$ as those
    whose commutator with the axial Killing vanishes at $Q$. Further
    results of the same type can be proven, as for instance the following
    two Lemmas which in some sense complete the previous one.
      
   \begin{lemma}
   For all vector fields $\vec{\alpha}$ and every point $Q\in W_2$, the vector 
$[\vec{\alpha},\vec{\sigma}] \mid_Q$ belongs to $P_Q$ and is orthogonal 
to $\vec{\alpha}\mid_Q$. 
   \end{lemma}
   {\it Proof :}
   Given that $\vec{\sigma}\mid_Q=0$ we have
\begin{equation}
    [\vec{\alpha},\vec{\sigma}]^{\beta}\mid_Q=\alpha^{\mu}\nabla_{\mu}
   \sigma^{\beta}\mid_Q \hspace{1cm} \forall \vec{\alpha}, \label{acovs}
\end{equation} 
   and also, as $\vec{\sigma}$ is a Killing vector,
    \[ [\vec{\alpha},\vec{\sigma}]^{\beta}\mid_Q=-\alpha^{\mu}\nabla^{\beta}
   \sigma_{\mu}\mid_Q. \]
   But then $[\vec{\alpha},\vec{\sigma}]^{\beta}V_{\beta}\mid_Q=0$ for all
   vectors $\vec{V} \in L_Q$, because if $\vec{V} \in L_Q$ then
   $V^{\beta}\nabla_{\beta}\sigma_{\mu}\mid_Q=0$ as follows from Lemma 3.
    Therefore $[\vec{\alpha},
   \vec{\sigma}]\mid_Q$ is orthogonal to every vector tangent to the
   axis and then $[\vec{\alpha},\vec{\sigma}]\mid_Q \ \in P_Q$. Finally,
    contracting
   (\ref{acovs}) with $\alpha_{\beta}$ it is obvious that $\vec{\alpha}
   \mid_Q$ and
   $[\vec{\alpha},\vec{\sigma}] \mid_Q$ are orthogonal to each other.
   \begin{lemma}
   $\vec{\alpha} \mid_Q$ is not tangent to the surface of fixed points at
   some point $ Q \in W_2 \\ ( \Leftrightarrow [\vec{\alpha},\vec{\sigma}]
   \mid_Q \neq \vec{0} $ already proven ) \ if and only if
   $[\vec{\alpha},\vec{\sigma}] \mid_Q $ is linearly independent of
   $\vec{\alpha} \mid_Q $ and $\vec{\sigma} \mid_Q$, and obviously then
   $[\vec{\alpha},\vec{\sigma}]$, as a vector field, is linearly independent
   of $\vec{\alpha}$ and $\vec{\sigma}$.
   \end{lemma} 
   {\it Proof :} $[\Leftarrow]$ \ If $[\vec{\alpha}.\vec{\sigma}]\mid_Q$
   is linearly independent of $\vec{\alpha}\mid_Q$ and $\vec{\sigma}\mid_Q$
   then $[\vec{\alpha},\vec{\sigma}]\mid_Q \not =\vec{0}$ and therefore,
   by Lemma 3, $\vec{\alpha}$ is not tangent to the axis.\\
   
   $[\Rightarrow]$  
    If $\vec{\alpha}\mid_Q$ is not tangent to the axis then we know by Lemmas 3
and 4 that $[\vec{\alpha},\vec{\sigma}]\mid_Q \ \in P_Q$, is different from
zero and orthogonal to $\vec{\alpha}\mid_Q$. Given then that
$[\vec{\alpha},\vec{\sigma}]\mid_Q$ is spacelike and that
$\vec{\sigma}\mid_Q=\vec{0}$ it follows that
$[\vec{\alpha},\vec{\sigma}]\mid_Q$ must be independent of
$\vec{\alpha}\mid_Q$ and $\vec{\sigma}\mid_Q$.

\vspace{0.5cm}

The combination of the three previous Lemmas gives rise to the following 
interesting theorem, which will be used repeatedly as a key point in
proving the main theorems of the next two sections.
   \begin{theorem}
   Let $\vec{\alpha}$ be a vector field in an axisymmetric space-time
   and $Q \in W_2$.
     \begin{enumerate}
     \item $\vec{\alpha} \mid_Q$ is tangent to the axis at $Q$ iff
     \ $[\vec{\alpha},\vec{\sigma}] \mid_Q=\vec{0}$.
     \item  $\vec{\alpha} \mid_Q (\neq \vec{0})$ is normal to the axis at $Q$
iff $\vec{\alpha} \mid_Q$ and $[\vec{\alpha},\vec{\sigma}] \mid_Q $ are
linearly independent vectors and $[[\vec{\alpha},\vec{\sigma}],\vec{\sigma}]
\mid_Q$ depends linearly on the previous.  \item  $\vec{\alpha}$ is neither
tangent nor normal to the axis at $Q$ iff $\vec{\alpha} \mid_Q, \
[\vec{\alpha},\vec{\sigma}] \mid_Q $ and
$[[\vec{\alpha},\vec{\sigma}],\vec{\sigma}] \mid_Q$ are linearly independent
vectors and $[[[\vec{\alpha},\vec{\sigma}],\vec{\sigma}], \vec{\sigma}] \mid_Q$
depends linearly on the previous two.  \end{enumerate} \end{theorem} {\it Proof
:}\\ Part 1) has already been shown. Let us go then to part 2).  If
$\vec{\alpha}\mid_Q \in P_Q$ , we have seen (Lemma 4) that
$[\vec{\alpha},\vec{\sigma}] \mid_Q \in P_Q$ and also that it is independent of
$\vec{\alpha}\mid_Q $. Thus $\vec{\alpha}\mid_Q$ and $[\vec{\alpha},
\vec{\sigma}]\mid_Q$ constitute an orthogonal basis for $P_Q$. But due to Lemma
4 again, $[[\vec{\alpha},\vec{\sigma}],\vec{\sigma}]\mid_Q \ \in P_Q$ and
therefore it is a linear combination of $\vec{\alpha}\mid_Q$ and
$[\vec{\alpha},\vec{\sigma}]\mid_Q$. The converse is shown in a similar manner.
   
   Finally, with regard to part 3),
   if $\vec{\alpha}\mid_Q \not\in P_Q$ and $\vec{\alpha} \mid_Q \not \in
   L_Q$ then we know, due to Lemma 4, first that
$[\vec{\alpha},\vec{\sigma}]\mid_Q \ \in P_Q$ and also that
$[[\vec{\alpha},\vec{\sigma}],\vec{\sigma}]\mid_Q \ \in P_Q$, and second, that
this last vector is orthogonal to $[\vec{\alpha},\vec{\sigma}]\mid_Q$.  As
$\vec{\alpha}\mid_Q \not \in P_Q$, then $\vec{\alpha}\mid_Q$,
$[\vec{\alpha},\vec{\sigma}]\mid_Q$ and $
[[\vec{\alpha},\vec{\sigma}],\vec{\sigma}]\mid_Q$ are linearly independent
vectors. Moreover, $[[[\vec{\alpha},
\vec{\sigma}],\vec{\sigma}],\vec{\sigma}]\mid_Q \ \in P_Q$ again by Lemma 4,
and then it must be a linear combination of $[\vec{\alpha},\vec{\sigma}]\mid_Q$
and $[[\vec{\alpha},\vec{\sigma}],\vec{\sigma}]\mid_Q$. The converse is now
obvious.

\vspace{0.3cm}

To end this section, we recall some interesting properties concerning the
intrinsic geometry of the axis of symmetry as well as the axial Killing vector
and its derivatives. Most of the following properties will also be used in
showing the main results of this paper presented below.

\vspace{0.3cm}

{\bf Property 1}. {\it The axis of symmetry is an autoparallel surface.}

 A surface in a manifold $V$ is autoparallel if for every pair of vector fields
defined on the surface and  tangent to the surface, say $\vec{X}$ and
$\vec{Y}$, the covariant derivative of $\vec{Y}$ along $\vec{X}$ remains
tangent to the surface. In our case, we have, for a point $Q \in W_2$,
$X^{\alpha}\nabla_{\alpha}Y^{\beta}\nabla
_{\beta}\sigma^{\mu}\mid_Q=X^{\alpha}\nabla_{\alpha}(Y^{\beta}\nabla_{\beta}
\sigma^{\mu})\mid_Q-X^{\alpha}Y^{\beta}\nabla_{\alpha}\nabla_{\beta}\sigma
^{\mu}\mid_Q$. Here the first term of the righthandside is zero because
$\vec{X}$ and $\vec{Y}$ are tangent to the axis and due to Lemma 3 and formula
(\ref{acovs}). The second term in the righthandside is zero too because being
$\vec{\sigma}$ a Killing vector field we can relate its second derivative with
the Riemann tensor of $V_4$ by $\nabla_{\alpha}\nabla_{\beta}
\sigma^{\mu}\mid_Q=\sigma_{\rho}{{R^{\rho}}_{\alpha\beta}}^{\mu}\mid_Q=0$.
Thus we have $X^{\alpha}\nabla_{\alpha}Y^{\beta}\nabla
_{\beta}\sigma^{\mu}\mid_Q=0$, which implies again that $X^{\alpha}\nabla
_{\alpha}Y^{\beta}$ is tangent to the axis and therefore that the surface $W_2$
is autoparallel.

\vspace{0.3cm}

 {\bf Property 2}. {\it The two second fundamental forms of the axis vanish
identically.}

 This is a general property of autoparallel surfaces (see e.g. \cite{SHO}).

\vspace{0.3cm}

{\bf Property 3}. {\it The tensor field
$\nabla_{\alpha}\sigma^{\rho}\nabla_{\beta} \sigma_{\rho}\equiv {\cal
H}_{\alpha\beta}$ is, at every point $Q\in W_2$, the projection tensor to
$P_Q$.}

 First of all, ${\cal H}_{\alpha\beta}$ is obviously a symmetric tensor.
 Moreover, for all vectors $\vec{V}$ tangent to the axis we have
 ${\cal H}_{\alpha\beta}V^{\beta}\mid_Q=0$ due to Lemma 3 and formula
(\ref{acovs}).  For non-zero vector fields orthogonal to the axis,
$\vec{P}\mid_Q \in P_Q$, we already know that $[\vec{P},\vec{\sigma}] \mid_Q$
belongs to $P_Q$ , is orthogonal to $\vec{P} \mid_Q$ and different from zero.
Moreover $P^{\alpha} {\cal
H}_{\alpha\beta}[\vec{P},\vec{\sigma}]^{\beta}\mid_Q=-P^{\alpha}
\nabla_{\alpha}\sigma^{\rho}\nabla_{\rho}\sigma_{\beta}P^{\mu}\nabla_{\mu}
\sigma^{\beta}\mid_Q=0$, where we have used twice the skew-symmetry of $
\nabla_{\rho}\sigma_{\beta}$.  So we have, $P^{\alpha}{\cal H}_{\alpha\beta}
\mid_Q=f(\vec{P},Q)P_{\beta}$ for every vector $\vec{P}\mid_Q \in P_Q$, where
$f$ is, in principle, a function depending on the point of the axis and the
vector $\vec{P} \mid_Q$.

 Let us now take another vector $\vec{R} \in P_Q$. We find $P^{\alpha} {\cal
H}_{\alpha\beta}\mid_QR^{\beta}=f(\vec{P},Q)P_{\beta}R^{\beta}\mid_Q=
f(\vec{R},Q) P_{\beta}R^{\beta}\mid_Q$ due to the symmetry of ${\cal H}_{\alpha
\beta}\mid_Q$ and then $f$ is a function depending only on the point in the
surface.  Taking two vectors fields on $W_2$, $\vec{u}$ and $\vec{v}$, which
are tangent to the axis and such that $u^{\alpha}u_{\alpha}=-1, \
v^{\alpha}v_{\alpha}=1$ and $u^{\alpha}v_{\alpha}=0$ everywhere on $W_2$, we
can write \[ {\cal H}_{\alpha\beta}\mid_Q=f(Q)(g_{\alpha\beta}
+u_{\alpha}u_{\beta}- v_{\alpha}v_{\beta})\mid_Q \] where $g$ is the metric in
$V_4$.  From the relation $\nabla_{\alpha}\nabla_{\beta}\sigma^{\mu}\mid_Q=0$
it is immediate to deduce that $\nabla_{\lambda}{\cal H}_{\alpha\beta}
\mid_{W_2}=0$ and from this expression one can easily find that $f$ is in fact
a constant rather than a function on $W_2$.  It only remains to prove that this
constant is equal to one. For a vector field such that $\vec{P}\mid_Q(\neq 0)
\in P_Q$  we have \[ P^{\alpha}{\cal
H}_{\alpha\beta}P^{\beta}\mid_Q=P^{\alpha}\nabla_{\alpha}
\sigma^{\rho}\nabla_{\beta}\sigma_{\rho}P^{\beta}\mid_Q=[\vec{P},\vec{\sigma}]
^{\rho}[\vec{P},\vec{\sigma}]_{\rho}\mid_Q=fP^{\rho}P_{\rho}\mid_Q\] and then f
is a positive constant, say $f=a^2$.

 It is a known result in the theory of transformation groups (see e.g.
 \cite{SHO}) that
 \[ d\tau_{\phi}(\vec{\alpha})=\vec{\alpha}+\phi[\vec{\alpha},\vec{\sigma}]
 +\frac{\phi^2}{2}[[\vec{\alpha},\vec{\sigma}],\vec{\sigma}]+ \dots \]
 In our case we have for the vector field $\vec{P}$, $[[\vec{P},
 \vec{\sigma}],\vec{\sigma}]^{\beta}\mid_Q= P^{\alpha}
 \nabla_{\alpha}\sigma^{\rho}\nabla_{\rho}\sigma^{\beta}\mid_Q=
 -P^{\alpha}{\cal H}^{\beta}_{\alpha}\mid_Q=-a^2P^{\beta}\mid_Q$, and therefore
it follows that, at $Q\in W_2$, \[ d\tau_{\phi}\mid_Q(\vec{P})=
\vec{P}\mid_Q+\phi[\vec{P},\vec{\sigma}]\mid_Q- \frac{\phi^2}{2}a^2
\vec{P}\mid_Q-\frac{\phi^3}{6}a^2[\vec{P},\vec{\sigma}]\mid_Q+ \dots = \] \[=
\cos(a\phi)\vec{P}\mid_Q+\frac{1}{a}\sin(a\phi)[\vec{P},\vec{\sigma}]\mid_Q \]
But if we now choose the standard parametrization for the torus such that
$\phi$ goes from $0$ to $2\pi$, then we have that
$d\tau_{2\pi}\mid_Q=Id\mid_Q$.  In consequence, and given that the realization
is effective, we have $a=1$ as required.

 Let us note finally that from the definition of ${\cal H}_{\alpha\beta}\mid_Q$
we can easily find the expression
\begin{equation}
{\cal H}_{\alpha\beta}\mid_Q=\frac{1}{2}
 \nabla_{\alpha}\nabla_{\beta}({\vec{\sigma}}^2)\mid_Q. \label{Hdds}
\end{equation}

\vspace{0.3cm}

{\bf Property 4}. {\it For any point $Q \in W_2$ there is a neighbourhood of
$Q$ such that ${\vec{\sigma}}^2$ is non-negative and it is zero only at points
on the axis}.

 This is trivial from the expression (\ref{Hdds}) just found in the previous
proof because ${\cal H}_{\alpha\beta}\mid_Q$ is positive definite and also both
${\vec{\sigma}}^2$ and $\nabla_{\rho}(\vec{\sigma}^2)$ vanish at the axis.
Therefore, there always exists a neighbourhood of the axis where the axial
Killing vector field is spacelike.

\vspace{0.3cm}

{\bf Property 5}. {\it At the axis of symmetry we have}
\[
\frac{\nabla_{\rho}(\vec{\sigma}^2)\nabla^{\rho}(\vec{\sigma}^2)}{4
\vec{\sigma}^2}
\longrightarrow 1. \]
This is the popular {\it regularity condition} of the axis of symmetry,
 which usually is
assumed to assure the standard $2\pi$-periodicity of $\phi$. We will prove this
result by using the projector ${\cal H}_{\alpha\beta}$ constructed above. Let
us evaluate $\nabla_{\alpha}{\vec{\sigma}}^2\nabla ^{\alpha}{\vec{\sigma}}^{2}=
4\sigma_{\rho}\sigma_{\nu}\nabla_{\alpha}
\sigma^{\nu}\nabla^{\alpha}\sigma^{\rho}=4\sigma^{\rho}\sigma^{\nu}
{\cal H}_{\rho\nu}$ near the axis up to second order, 
by means of an expansion in a coordinate system.
At any point with coordinates $x^{\nu}$, in some neighbourhood of
$Q \in W_2$ with coordinates $Q^{\nu}$, we obviously have
\[ \sigma^{\mu}(x)=(x^{\nu}-Q^{\nu})\nabla_{\nu}\sigma^{\mu}\mid_Q+o(2)\]
\[ {\cal H}_{\mu\nu}(x)={\cal H}_{\mu\nu}\mid_Q+o(1)\]
and then we find
$ \sigma^{\mu}\sigma^{\nu}{\cal H}_{\mu\nu}(x)=(x^{\alpha}-Q^{\alpha})
\nabla_{\alpha}\sigma^{\mu}(x^{\beta}-Q^{\beta})\nabla_{\beta}\sigma^\nu
\nabla_{\mu}\sigma^{\rho}\nabla_{\nu}\sigma_{\rho}\mid_Q+o(3)= \\ (x^{\alpha}-
Q^{\alpha})(x^{\beta}-Q^{\beta}){{\cal H}_{\alpha}}^{\rho}{\cal H}_{\beta
\rho}+o(3)=(x^{\alpha}-Q^{\alpha})(x^{\beta}-Q^{\beta}){\cal H}_{\alpha\beta}
\mid_Q+o(3).$\\
On the other hand from the expansion of $\sigma^{\mu}(x)$ we have
\begin{eqnarray}
{\vec{\sigma}}^2(x)=(x^{\alpha}-Q^{\alpha})(x^{\beta}-Q^{\beta})
\nabla_{\alpha}\sigma_{\rho}\nabla_{\beta}\sigma^{\rho}\mid_Q+o(3)=
\nonumber \\
(x^{\alpha}-Q^{\alpha})(x^{\beta}-Q^{\beta}){\cal H}_{\alpha\beta}\mid_Q+o(3)
\nonumber
\end{eqnarray}
From this expression follows immediatly the property above. Moreover, in a
linearized point of view of the manifold, and in a neighbourhood of $Q \in
W_2$, we can consider $(x^{\alpha}-Q^{\alpha})$ as a vector pointing from $Q$
to $x$.  Its double contraction with ${\cal H}_{\alpha\beta}$ gives the modulus
of the normal component of this vector in $P_Q$ or, in other words, the
distance from the axis to the point $x$. The last expression shows that the
norm ${\vec{\sigma}}^2\mid_x$ coincides, at first non-trivial order, with this
distance from $x$ to the axis of symmetry. If we now consider the orbit of the
axial symmetry which passes through $x$ we have that its length is
$2\pi{\vec{\sigma}}^2\mid_x$, because the norm of a Killing vector field
remains constant alongs its orbits. Then we have found that the length of the
axial orbit is, at first relevant order, $2\pi$ times the distance from $x$ to
the axis of symmetry, which states what is ususally called {\it elementary
flatness}.

\section{Axially symmetric spacetimes with conformal symmetries.}

The results of the previous section hold in space-times with axial symmetry. 
However, we are interested in $V_4$'s which also have conformal symmetries.
This section is devoted to presenting the basic results appearing when these
two symmetries hold simultaneously, which are the main results in this paper.

Let us start by recalling that a conformal isometry of $V_4$ is a
diffeomorphism
\begin{eqnarray*}
 L : V_4 &\rightarrow& V_4 \\
 x &\rightarrow& L(x) 
\end{eqnarray*}
such that $L^{\star}g=e^Ug$, where $L^{\star}$ is the pullback 
of $L$, $g$ is the metric tensor field in $V_4$ and $e^U$
is a function called the conformal factor of $L$.
We consider the case where there is a one-dimensional group of transformations
in $V_4$, say $\symbol{123}\Lambda_s\symbol{125}$, which are conformal
isometries with associated conformal Killing vector field $\vec{\lambda}$. That
is to say, there is a map
\begin{eqnarray*}
      \Lambda  :  \Re \ (or \hspace{0.1cm} T) \ \times V_4 &\rightarrow& V_4 \\
(s,x) &\rightarrow& \Lambda(s,x) \equiv \Lambda_s(x)
\end{eqnarray*}
which is a realization of the one-dimensional Lie group $\Re$ (or $T$) in
$V_4$, where each $\Lambda_s$ is a conformal isometry of the spacetime:
\begin{equation}
\Lambda_s^{\star}g=e^{U_s}g. \label{Us}
\end{equation}
The infinitesimal generator of this group of transformations, $\vec{\lambda}$,
satisfies then
\begin{equation}
      {\cal L}_{\vec{\lambda}} g = \Psi g \label{kc}
\end{equation}
where ${\cal L}_{\vec{\lambda}}$ is the Lie 
derivative with respect to $\vec{\lambda}$ and $\Psi(x)$ is a function called 
the scale factor of $\vec{\lambda}$. It is easily verified that
\begin{equation}
 U_s(x) = \int ^{s}_{0}
    (\Psi\circ\Lambda_h(x))dh \Longleftrightarrow \Psi (x)=\frac{d}{ds}U_s(x)
\mid_{s=0}. \label{psi}
\end{equation}

We begin by proving the following general result that we will need later:
    \begin{p}
    Let $L$ be a conformal transformation in $V_4$ with conformal
    factor $e^U$ and $\vec{\xi}$ a Killing vector field. The vector field
    defined by \[ \vec{\zeta}(y) = dL(\vec{\xi}(x))  \ \ \
    y = L(x) \] is a conformal Killing field and it is a Killing
    field if and only if $\vec{\xi}(U) = 0$.
    \end{p}
    {\it Proof :}  We have $L^{\star}g = e^{U}g$ where $U$ is a
scalar function, and $\Xi_{s}^{\star}g = g$ where $\{\Xi_s\}$ is the
local group of transformations generated by $\vec{\xi}$.
    
    Of course $L^{-1}$ is a conformal transformation and
it is easy to check that \[ {L^{-1}}^{\star}(g)= e^{-(U
\circ L^{-1})} g \] Let us now define $ L_s
=L\circ\Xi_s\circ L^{-1} $.  Obviously $\{L_s\}$ is
a local one-parameter group of conformal isometries. In fact 
\[ L_s^{\star}g \mid_x =
{(L\circ \Xi_s\circ L ^{-1})}^{\star}(g)\mid_x =
{(L^{-1})}^{\star}\circ{(\Xi_s)}^
{\star}\circ{(L)}^{\star} (g)\mid_x =
(L^{{-1}^{\star}}\circ \ {\Xi_s}^{\star})(e^{U}g)\mid_x \]
\[={L^{-1}}^{\star}(e^{U\circ\Xi_s}g) \mid_x=
e^{U\circ\Xi_s \circ L^{-1}}e^{-U\circ L^{-1}} g\mid_x =
e^{U\circ\Xi_s\circ L^{-1}-U\circ L^{-1}}g \mid_x. \]
A very well-known standard fact (see, for example \cite{CHO}) is that the
infinitesimal generator of $\{L_s\}$ is given by $\vec{\zeta}=dL(\vec{\xi})$.
Thus, we have that $dL(\vec{\xi})$ is a conformal Killing vector field. It will
actually be a Killing vector field if and only if, for every value of $s$
\hspace{0.2cm} $U\circ\Xi_s\circ L^{-1} - U\circ L^{-1} = 0$, that is to say,
iff $U(\Xi_s(x))=U(x)$ which means that  $U$ is constant along the orbits of
$\vec{\xi}$. So $dL(\vec{\xi})$ is a Killing vector field if and only if
$\vec{\xi}(U)=0$, as we wanted to prove.

\vspace{0.2cm}
      
    The same result can be essentially deduced by using the expression
    \[ {\cal L}_{\vec{\xi}}(L^{\star} g)=L^{\star}({\cal L}_{
    dL(\vec{\xi})} g) \]
    that can be shown for all vectors $\vec{\xi}$, all covariant tensors $g$
and all $L : V_4 \rightarrow V_4 $.

    \vspace{3mm}
    
    {\bf Corollary} {\it Let $\{\Lambda_s\}$ be a one-dimensional
    group of conformal transformations
    in $V_4$ with scale factor $\Psi$ and $\vec{\xi}$ a Killing vector field.
    For each $s$, the vector field defined by
    \[ \vec{\zeta}_s(y)=d\Lambda_s(\vec{\xi}(x)) \ \ \ y= \Lambda_s(x) \]
    is a conformal Killing vector field and it is a Killing vector field iff
$\vec{\xi} (\Psi)=0$.}
    
    \vspace{3mm}
    
    The corollary follows because from the previous Lemma the condition is
$\vec{\xi}(U_s)=0$ and then from formula (\ref{psi}) this is equivalent to
$\vec{\xi}(\Psi)=0$.

\vspace{0.2cm}

We can now come back to the case in which the spacetime is axisymmetric. The
notations are then those of section 2. The first interesting fact is: \begin{p}
In an axisymmetric spacetime, let $\vec{\lambda}$ be a conformal Killing vector
field tangent to the axis for all $Q \in W_2$ and with associated scale factor
$\Psi$.  Then $[\vec{\lambda},\vec{\sigma}] = \vec{0}$ if and only if
$\vec{\sigma} (\psi) = 0.$ \end{p}

   {\it Proof :}
   For every $s$ we define the following realization of $T$ into $V_4$
\begin{eqnarray*}
   \tilde{\tau}^{(s)} : T \times V_4 &\rightarrow& V_4 \\
    (\phi,x) &\rightarrow& \tilde{\tau}^{(s)}(\phi,x)\equiv\tilde{\tau}
   _{\phi}^{(s)}(x)
\end{eqnarray*}
   by \( \tilde{\tau}_{\phi}^{(s)}=\Lambda_s\circ\tau_{\phi}\circ\Lambda
   _{-s} \)
   where $\{\Lambda_s\}$ is the local one-parameter group of transformations
   generated by $\vec{\lambda}$. Given that $\vec{\lambda}$ 
   is tangent to the axis, it is straightforward to see that the fixed points
of $\tilde{\tau}_{\phi}^{(s)}$ are the fixed points of $\tau_{\phi}$, and
viceversa.  We start by proving first the converse implication of the
Proposition.
 
   $[\Leftarrow]$ If $\vec{\sigma}(\Psi)=0$ then $\tilde{\tau}_{\phi}^{(s)}$ is
an isometry  for each $\phi$ (because of Proposition 1 and its corollary) which
is a realization of the Lie group $T$ and with the same axis of symmetry that
$\tau$. A Lemma due to Carter \cite{C} implies that they are the same group of
transformations. So we have for every $s$ and $\phi$,
$\Lambda_s\circ\tau_{\phi}\circ\Lambda_{-s}=\tau_{\phi}$, which implies (see
e.g. \cite{CHO}) $[\vec{\lambda},\vec{\sigma}]=\vec{0}$.
  
  $[\Rightarrow]$ If $[\vec{\lambda},\vec{\sigma}]=\vec{0}$, then we have 
  \cite{CHO} $\Lambda_s\circ\tau_{\phi}\circ\Lambda_{-s}=\tau_{\phi}$ which
means that $\tilde{\tau}_{\phi}^{(s)}$ is an isometry. It follows from Lemma 6
and its corollary that $\vec{\sigma}(\Psi)=0 $.

\vspace{0.2cm}

  Therefore, the necessary and sufficient condition such that a conformal
  Killing vector field tangent to the axis commutes with the axial Killing
  is that the scale factor be constant along the orbits of the axial
  Killing vector field . A trivial consequence is that all homothetic Killing 
  vector fields (and also all Killing vector fields) tangent to the axis 
  commute with the axial
  symmetry. Despite of this, it seems that, in principle, this is not true
  for general conformal Killing vector fields. We shall see, however, that this
  property does hold for general conformal Killings. In order to prove
  it, we first need the following fundamental result, which strengthes
  previous similar results \cite{C} and states that an axial conformal Killing
vector field and an axial Killing vector field with the same axis of symmetry
in a given spacetime must coincide.
    
    \begin{theorem}
    Let $\{\eta_{\phi}\}$ be an effective  realization of the Lie group $T$
that is a conformal isometry with a surface of fixed points $\tilde{W_{2}}$ in
an axisymmetric space-time. If $\tilde{W_2} = W_2$, then $\{\eta_{\phi}\}$ is
in fact an isometry and coincides with the axial isometry.  \end{theorem} {\it
Proof :} We call $\vec{\mu}$ the infinitesimal generator of $\{\eta_{\phi}\}$.
The method of the proof is to see that both $\vec{\mu}$ and $\vec{\sigma}$
satisfy the same equations and the same initial conditions, so that they must
be identical. The vector field $\vec{\mu}$ is a conformal Killing and then it
satisfies
\begin{equation}
    {\cal L}_{\vec{\mu}}(g)=\tilde{\Psi}g \label{kct}
\end{equation}
where $\tilde{\Psi}$ is the associated scale factor. On the other hand,
    it follows from (\ref{psi}) that for conformal Killing vector fields the
finite transformation verifies
\begin{eqnarray*}
 \eta_{\phi}^{\star}(g)(x)=\exp \left( \int^{\phi}_{0}(\tilde{\Psi}\circ\eta_
    {\rho})(x)d\rho \right) \ g(x) .
\end{eqnarray*}
   
 Let $Q$ be a point on the axis of symmetry $W_2$ of $\symbol{123}
    \eta_{\phi}\symbol{125}$. As $\eta_{2\pi}=Id\mid_{V_4}$, 
    we then have $\int_{0}^{2\pi}(\tilde{\Psi}\circ
    \eta_{\rho})(Q) d\rho = 0$ .
    But $Q$ is a fixed point and thus, for every $\rho \in T$,
$\eta_{\rho}(Q)=Q$ so that from the previous equation we obtain that the scale
factor vanishes at the axis:
\begin{equation}
 \tilde{\Psi}\mid_{W_2}=0. \label{psi0}
\end{equation}
    
    Analogously to Lemma 1, the vectors tangent
    to the axis of symmetry in a fixed point can be characterized by
    \[ \vec{V} \in L_Q \Leftrightarrow d\eta_{\phi}\mid_Q(\vec{V})=\vec{V}
\hspace{0.4cm} \forall \phi \in T \]
    
    Our aim, now, is to see that for a point $Q\in W_2$, the differential
    maps $d\tau_{\phi}\mid_Q$ and $d\eta_{\phi}\mid_Q$ coincide. First,
    $d\eta_{\phi} \mid_Q$ is an automorphism of $T_Q (V_4)$ and leaves
    $L_Q$ invariant. Moreover it is an isometry of $T_Q (V_4)$ because
    $\tilde{\Psi}(Q) = 0 $. In consequence $P_Q$, the complement orthogonal
    of $L_Q$ in $T_Q (V_4)$, is an invariant subspace of $d\eta_{\phi}\mid_Q$.
Thus,  we have two effective isometric realizations of the Lie group $T$ on
$P_Q$ , which is a two-dimensional vector space with positive definite metric,
and then they must coincide. So we have for every point $Q$ on the axis
$d\tau_{\phi}\mid_Q=d\eta_{\phi}\mid_Q$, from where it can be immediately
deduced that
\begin{equation} 
    \nabla \vec{\mu} \mid_Q=\nabla\vec{\sigma}\mid_Q  \label{fd}
\end{equation}
where we have used 
\begin{equation}
\vec{\sigma}\mid_Q =\vec{\mu}\mid_Q =\vec{0}. \label{sm}
\end{equation}   
    
From formula (\ref{fd}) it follows that for every vector $\vec{V}\in L_Q$ we
have
\begin{equation}
       V^{\alpha}\nabla_{\alpha}
    (\nabla_{\beta}\mu_{\gamma})\mid_Q=V^{\alpha}\nabla
    _{\alpha}(\nabla_{\beta}\sigma_{\gamma})\mid_Q \label{s}
\end{equation}
    But, as $\vec{\mu}$ is a conformal Killing vector field and $\vec{\sigma}$
is a Killing vector field, they obey the following relations, see e.g.
\cite{EIS}
\begin{eqnarray}
     \nabla_{\rho}\nabla_{\nu}\mu_{\alpha}&=&\mu_{\delta}
    R_{\rho\nu\alpha}^{\delta} +
\frac{1}{2}(g_{\nu\alpha}\nabla_{\rho}\tilde{\Psi} +g_{\alpha\rho}
\nabla_{\nu}\tilde{\Psi}  - g_{\rho\nu}\nabla_{\alpha}\tilde{\Psi}),
\label{ddm} \\ \nabla_{\rho}\nabla_{\nu}\sigma_{\alpha}&=& \sigma_
{\delta}R_{\rho\nu\alpha}^{\delta} \nonumber
\end{eqnarray}
    and this together with (\ref{sm}) gives
\begin{eqnarray}
    \nabla_{\rho}\nabla_{\nu}\mu_{\alpha}\mid_Q&=&
    \frac{1}{2}(g_{\nu\alpha}\nabla_{\rho}\tilde{\Psi}
     +g_{\alpha\rho} \nabla_{\nu}\tilde{\Psi}  -
    g_{\rho\nu}\nabla_{\alpha}\tilde{\Psi})\mid_Q, \label{ss} \\    
    \nabla_{\rho}\nabla_{\nu}\sigma_{\alpha}\mid_Q&=&0. \label{sss}
\end{eqnarray}
    Combining (\ref{s}) with (\ref{sss}) we can write, for every $\vec{V} \in
L_Q$ \[ V^{\alpha}\nabla_{\alpha}
(\nabla_{\beta}\mu_{\gamma})\mid_Q=V^{\alpha}\nabla
_{\alpha}(\nabla_{\beta}\sigma_{\gamma})\mid_Q=0  \] and this last expression
together with (\ref{ss}) leads us to
\begin{equation}
    \frac{1}{2}(V_{\alpha}\nabla_{\nu} 
    \tilde{\Psi}-V_{\nu}\nabla_{\alpha}\tilde{\Psi})\mid_Q=0 \label{ssss}
\end{equation}
    where we have used (see formula (\ref{psi})) that
$V^{\rho}\nabla_{\rho}\tilde{\Psi}\mid_Q=0$, that is to say,
$\nabla_{\alpha}\tilde{\Psi}\mid_Q \in P_Q$.  But then, being $\vec{V} \in L_Q$
and $\nabla_{\alpha}\tilde{\Psi} \in P_Q$, expression (\ref{ssss}) implies
necessarily
\begin{equation} 
    \nabla_{\alpha}\tilde{\Psi}\mid_Q=0. \label{dpsi}
\end{equation}

    Equations (\ref{kct}) for a conformal Killing vector field $\vec{\mu}$ with
scale factor $\tilde{\Psi}$ are
\begin{equation}
    \nabla_{\rho}\mu_{\delta}+\nabla_{\delta}\mu_{\rho}= \tilde{\Psi} \ g
    _{\rho\delta}. \label{kct2}
\end{equation}   
    These equations are not written in normal form, but well-known
    consequences of them \cite{EIS} are expression (\ref{ddm}) and the
following
\begin{equation}
    \nabla_{\beta}\nabla_{\alpha}\tilde{\Psi}= \frac{1}{6}({\cal L}_{\vec{\mu}}
    R+ \tilde{\Psi}R)g_{\beta\alpha}-{\cal L}_{\vec{\mu}}R_{\beta\alpha},
\label{ddpsi}
\end{equation}
where $R_{\beta\alpha}$ and $R$ are the Ricci tensor and the scalar curvature
of the spacetime, respectively.

    Expressions (\ref{kct2}), (\ref{ddm}) and (\ref{ddpsi}) imply, among other
things, that if at some point $x \in V_4$ a conformal Killing vector field and
its scale factor have the following properties \[ \vec{\mu}\mid_x= \vec{0} \ \
\ \ , \ \nabla_{\nu}\mu_{\rho}\mid_x=0\] \[ \tilde{\Psi}(x)=0 \ \ \ \ , \
\nabla_{\nu}\tilde{\Psi}\mid_x=0\] then the conformal Killing vector field must
be zero everywhere.
    
    In our case, the vector field $\vec{\mu}-\vec{\sigma}$ is
    a conformal Killing with scale factor $\tilde{\Psi}$
    and with the properties written above, because of formulae 
    (\ref{psi}), (\ref{dpsi}), (\ref{sm}) and (\ref{fd}).
    Consequently, $\vec{\mu}-\vec{\sigma}$ must be zero
    everywhere. Therefore, $\vec{\mu}$ is a Killing vector field and the
    conformal symmetry it defines is, in fact, the axial symmetry we already
had in the spacetime.

\vspace{0.3cm}
    
    We are now prepared to prove the main theorems in this paper. Let us
    start with the following important result
    
    \begin{theorem}
    In an axially symmetric spacetime, if $\vec{\lambda}$ is a conformal
Killing vector field tangent to the axis of symmetry for all $Q\in W_2$, then
{\large \[ [\vec{\lambda},\vec{\sigma}]=\vec{0} \]} and also Proposition 2
implies that $\vec{\sigma}(\Psi)=0$, where $\Psi$ is the scale factor
associated with $\vec{\lambda}$.  \end{theorem} {\it Proof :} Let us call, as
before, $\{\Lambda_s\}$ the local group of transformations generated by
$\vec{\lambda}$ with $s$ in some neighbourhood of zero and
$\tilde{\tau}_{\phi}^{(s)}=\Lambda_s\circ\tau_{\phi}\circ\Lambda_{-s}$.  We
have already shown (proof of Proposition 2) that the set of fixed points of $
\tilde{\tau}^{(s)}$ are the same that the set of fixed points of $\tau$.  Then
we can apply the previuos theorem to see that, for every $\phi$ and $s$, \[
\tilde{\tau}_{\phi}^{(s)}  = \Lambda_s\circ\tau_{\phi}\circ\Lambda_{-s} =
\tau_{\phi} \] and consequently, we get \[
[\vec{\lambda},\vec{\sigma}]=\vec{0}. \]

\vspace{0.5cm}

    This result allows us to prove that in axially symmetric spacetimes with
    a conformal Killing and no other conformal Killing (nor Killing)
    vector fields, the axial symmetry and the
    conformal symmetry commute. More precisely we have the following theorem.
    \begin{theorem}
    In an axisymmetric spacetime with a conformal Killing vector
    $\vec{\lambda}$, if there is no more conformal symmetry then
    {\large \[ [\vec{\lambda},\vec{\sigma}]=\vec{0}. \] }
    \end{theorem}
    {\it Proof :}    
    For a point $Q \in W_2$, if we had
$[\vec{\sigma},\vec{\lambda}]\mid_Q\not=\vec{0}$ then (Theorem 2)
$[\vec{\sigma},\vec{\lambda}]\mid_Q$ would be linearly independent of
$\vec{\lambda}\mid_Q$ and $\vec{\sigma}\mid_Q$, and then as vector fields they
are independent. As $[\vec{\sigma},\vec{\lambda}]$ is a conformal Killing
vector field we would have more symmetry against the hypothesis.  So, for every
$Q\in W_2$,  $[\vec{\sigma},\vec{\lambda}] \mid_Q=\vec{0}  $ and then, because
of the same Theorem 2, $\vec{\lambda}$ is tangent to the surface of fixed
points for all $Q\in W_2$. Theorem 4 implies then that
$[\vec{\sigma},\vec{\lambda}]=\vec{0}$.

\vspace{0.5cm}

    Another immediate consequence of the previous results is that, in an
axially symmetric spacetime, there is no restriction in assuming that a
timelike conformal Killing commutes with the axial Killing (this was already
known for Killing vector fields, see \cite{C}). The precise statement is given
in the corollary following the next Proposition.
\begin{p} 
In an axisymmetric spacetime, let $\vec{\lambda}$ be a conformal Killing
vector field which does not commute with $\vec{\sigma}$.
If at some point $Q$ of the axis $\vec{\lambda}\mid_Q$ is not normal to the 
surface of fixed points, then there always exists another conformal Killing
vector field that commutes with the axial Killing vector field.
\end{p}
{\it Proof :}
We know that $\vec{\lambda}+[[\vec{\lambda},\vec{\sigma}],\vec{\sigma}]$ is a
conformal Killing vector field. We are going to see that this vector field is
tangent to the axis everywhere on the axis. In fact, at any point $Q \in W_2$,
we can decompose $\vec{\lambda}\mid_Q$ in an unique way into its components
tangent and normal to the surface:
$\vec{\lambda}\mid_Q=\vec{\lambda}_{\|}\mid_Q+\vec{\lambda}_{\bot}\mid_Q$.  On
the other hand, we have
$[[\vec{\lambda},\vec{\sigma}],\vec{\sigma}]\mid_Q=-\vec{\lambda}_{\bot}\mid_Q$
due to property 3 of axial symmetry listed at the end of section 2.  In
consequence, the conformal Killing vector field
$\vec{\lambda}+[[\vec{\lambda},\vec{\sigma}],\vec{\sigma}]$ is tangent to the
axis at every point $Q \in W_2$ and then, by Theorem 4, it commutes with the
axial Killing vector field. If $\vec{\lambda}$ is not orthogonal to the axis at
some point $Q \in W_2$ then the conformal Killing vector field considered is
not identically vanishing.

\vspace{3mm}

{\bf Corollary} {\it In an axisymmetric spacetime, if there is a timelike
conformal Killing field, 
then there always exists a timelike conformal Killing field that commutes with
the axial Killing field.}

\vspace{3mm}

The corollary is evident because a timelike vector field cannot be orthogonal
to the axis anywhere and
its tangent component is obviously timelike because of Theorem 1. Thus, 
the derived conformal Killing
field of the previous proposition commutes with tha axial symmetry and is also
timelike
at least in the region where the original one was timelike, as can be checked.

Let us remark that all the results shown in this section for conformal
Killings hold also for homothetic Killings and real Killings, as is obvious.
Most of these results were known for Killing fields but, as far as we
know, they were previously unknown for general conformal Killing vector fields.

\section{Axisymmetric spacetimes with another symmetry and a conformal
symmetry.}
       
    Until now we have been considering an axisymmetric spacetime with a
    conformal Killing vector field.  In General Relativity it has much
    interest the case of axisymmetric spacetimes with another symmetry
    which commutes with the axial symmetry, for example stationary and
    axisymmetric spacetimes or cylindrically symmetric spacetimes.
    It is obvious that all we have done in the case of conformal 
    Killing fields applies for Killing fields as well, and so we can
    recover the main results proved by Carter in the early
    seventies. Moreover, it has been recently found \cite{K3} that a class
    of stationary and axisymmetric exact solutions \cite{S2} possesses a
    conformal Killing vector and a new family of stationary and axisymmetric
exact solution with the same property has been constructed \cite{S3}. Due to
the importance that these type of metrics may have in describing the
gravitational field of isolated objects, as explained in the Introduction, some
papers have recently appeared in the literature considering the case of
stationary and axisymmetric exact solutions with a third proper conformal
Killing vector field and studying the different Bianchi types that these three
vector fields can adopt.
    
    In that direction the previous Lemmas and Theorems allow us to show the
following result.
    
    \begin{lemma} In an axisymmetric spacetime with another Killing vector
field $\vec{\xi}$ and a proper (or homothetic) conformal Killing field
$\vec{\lambda}$, if there is no more conformal symmetry then \[
[\vec{\sigma},\vec{\xi}]=\vec{0} \] and also $\vec{\lambda}\mid_Q$ is tangent
to the axis for all $Q\in W_2$.  \end{lemma} Here no more conformal symmetry
means no more Killing vector fields and no more conformal Killing vector fields
either.
    
    {\it Proof :} The first part of this proof follows exactly the same steps
that those in the proof of Theorem 5, because the results proven for conformal
Killing vector fields are also true for Killing vector fields and we are
assuming that there are no Killing fields other than $\vec{\sigma}$ and
$\vec{\xi}$.
    
    With respect to the second part, we know that
$[\vec{\lambda},\vec{\sigma}]$ is a conformal Killing vector field, so that the
only possibility for not having more conformal symmetry is that
\begin{equation}
 [\vec{\lambda},\vec{\sigma}]= a\vec{\sigma} + b\vec{\lambda} + c\vec{\xi}
\label{last}
\end{equation}
where $a,b$ and $c$ are constants.  If we commute now with $\vec{\sigma}$ we
obtain $[[\vec{\lambda},\vec{\sigma}],\vec{\sigma}]=b
[\vec{\lambda},\vec{\sigma}]$.  But at points of the axis we know by Lemma 4
that $[[\vec{\lambda},\vec{\sigma}],\vec{\sigma}]\mid_{W_2}$ and
$[\vec{\lambda},\vec{\sigma}]\mid_{W_2}$ are normal to the axis and orthogonal
to each other. It must be then,
$b[\vec{\lambda},\vec{\sigma}]\mid_{W_2}=\vec{0}$.  If
$[\vec{\lambda},\vec{\sigma}]\mid_{W_2}=\vec{0}$ we are done because of Lemma
3.  Otherwise we should have $b=0$, and then from (\ref{last}) and given that
$\vec{\xi}\mid_{W_2}$ is tangent to the axis we obtain again by Lemma 4 that
$[\vec{\lambda},\vec{\sigma}]\mid_{W_2}=\vec{0}$, and using Lemma 3,
$\vec{\lambda}\mid_{W_2}$ is tangent to the axis.

\vspace{0.3cm}
    
    Trivial consequence of this Lemma and Theorem 4 is the following important
result.  \begin{theorem} In an axisymmetric spacetime with another Killing
vector field $\vec {\xi}$ and a conformal Killing vector field $\vec{\lambda}$,
if there is no more conformal symmetry then

    {\large \[ [\vec{\sigma},\vec{\xi}]=\vec{0}, \] \[
[\vec{\sigma},\vec{\lambda}]=\vec{0}. \]      } \end{theorem}

We see, therefore, that in stationary and axisymmetric spacetimes, if there is
one (and only one) proper (or homothetic) conformal Killing vector field, then
it must commute with the axial symmetry. Of course, the same happens in
cylindrically symmetric spacetimes.  This is a very interesting result and, in
fact, it simplifies largely the Bianchi types one has to study in these cases.
Thus, for instance, it has been recently considered the case of stationary and
axisymmetric perfect-fluid spacetimes in Refs.\cite{KC}, \cite{C-AC}. In these
papers, Bianchi types with $[\vec{\sigma},\vec{\lambda}]\neq \vec{0}$ have been
studied with the result of the impossibility of getting solutions to the field
equations for a perfect-fluid energy-momentum tensor. In fact, as we have
shown, it does not matter what the energy-momentum is, there exists {\it no}
spacetime with that property. Theorem 6 above should be taken into account for
future work in spacetimes with two symmetries and one conformal symmetry,
whenever one of the symmetries is required to be axial. Similarly, for cases
with axial symmetry and only one more conformal symmetry we have proven in the
previous section that these two symmetries must commute. Therefore, if we want
to study conformally stationary and axially symmetric spacetimes, we can assume
without restriction that these two symmetries commute (analogously to what
happens in stationary and axisymmetric manifolds), and set up the coordinate
system accordingly. Many other consequences can be extracted from the results
herein shown, but as they are self-evident we do not believe necessary to
explain them here further.

\end{document}